\documentclass{article} 
\usepackage{iclr2025_conference,times}

\usepackage{amsmath,amsfonts,bm}

\def\eqref#1{equation~\ref{#1}}

\def\1{\bm{1}}

\DeclareMathAlphabet{\mathsfit}{\encodingdefault}{\sfdefault}{m}{sl}
\SetMathAlphabet{\mathsfit}{bold}{\encodingdefault}{\sfdefault}{bx}{n}

\usepackage{enumitem}
\usepackage{hyperref}
\usepackage{url}
\usepackage[most]{tcolorbox}
\usepackage{xcolor}
\usepackage{colortbl}
\usepackage{algorithm}
\usepackage{algpseudocode}

\title{A Note on Code Quality Score: LLMs for Maintainable Large Codebases
}
\author{Sherman Wong\thanks{Equal contribution. Please email to\texttt{ shermanwong@meta.com, jalajbhandari@meta.com}}, Jalaj Bhandari$^{*}$, Leo Zhou Fan Yang, Xylan Xu, Yi Zhuang, \\
\textbf{Cem Cayiroglu, Payal Bhuptani, Sheela Yadawad \& Hung Duong} \\
Meta Platforms, Inc.
}

\newcolumntype{P}[1]{>{\centering\arraybackslash}p{#1}}
\newcolumntype{C}[1]{>{\centering\arraybackslash}m{#1}}
\usepackage{array}
\usepackage{makecell}

\iclrfinalcopy 
\begin{document}

\maketitle

\begin{abstract}
Maintaining code quality in large-scale software systems presents significant challenges, particularly in settings where a large numbers of engineers work concurrently on a codebase. This paper introduces Code Quality Score (CQS) system to automatically detect issues with a set of code changes and provide actionable insights. At its core, the CQS system is powered by two Llama3 models, fine-tuned (with SFT and offline RL approaches), to a) detect common code quality issues related to coding best practices and b) to provide good ``critiques'' for LLM-generated code review respectively. To maintain good user experience, we layer the system with hand-crafted rules to filter out incorrect responses/hallucinations. Offline evaluations show that our CQS system is able to achieve an impressive precision rate for identifying valid issues. This system has already been rolled out to developers in an industrial scale setting and has consistently achieved 60\% week over week user helpfulness rate, demonstrating its effectiveness in a real-world environment. In this paper, we present details of the CQS system along with some learnings on curating developer feedback to create training data for LLM fine-tuning.

\end{abstract}

\section{Introduction}
Evaluating code quality is a crucial part of the software development cycle, in open-source and industrial settings, that requires a significant amount of developer time and effort \citep{sadowski2018modern}. Recent studies point to high code quality as one of the key drivers of increased productivity of software engineering organizations and can have a huge impact on the number of defects, issue resolution time as well as time to development \cite{cheng2022improves,tornhill2022code}. Challenges to maintain a high quality codebase tend to scale with size and complexity in large industrial settings, with many engineers working concurrently on a continuously evolving codebase and issues related to high cyclomatic complexity, legacy code, inadequate documentation etc. requiring resources to manage the technical debt. 

However, despite increasing awareness about its importance, the process of evaluating code quality remains mostly manual and is typically done via peer code review. While traditional static analysis tools like linters are often used to automatically check some aspects of code quality such as formatting, other important (and often nuanced) aspects of code quality such as those related to code modularity, readability, error handling, testability etc. are ambiguous and hard to evaluate by such rule based systems. This, along with high developer satisfaction rates \citep{winters2020software}, has led to human code review as the dominating paradigm.

Advances in deep learning and natural language processing (NLP) have inspired initial attempts to automate the code review process with large scale pre-trained models, focusing on specific tasks such as code quality estimation, review generation and code refinement \citet{li2022automating, thongtanunam2022autotransform, tufano2022using, li2022auger}. Most of these ideas require domain and task specific pre-training necessitating substantial computational resources. Large language models (LLMs) on the other hand have shown impressive cross domain generalization across a variety of tasks without the need for specific pre-training -- in fact, most of the recent advances have leveraged post-training methods to further improve LLM performance on specific tasks. 

We take inspiration from some very recent work which demonstrate early success of using LLMs for automated code review \citep{lu2023llama, vijayvergiya2024ai,coderabbitai2024awesome} and develop a Code Quality Score (CQS) system -- an LLM-powered system for code quality evaluation and code review at industrial scale settings. We break down the entire code review process into three distinct tasks, namely, ``issue collection'', ``issue validation'' and ``action generation'' and 
design CQS to orchestrate multiple components with each component responsible for one of these tasks. For different CQS components, we fine-tune state-of-the-art open source LLMs (Llama models) with post-training methods like supervised fine-tuning (SFT) and Direct Preference Optimization (DPO), an offline RL approach. CQS has already been deployed to a large number of developers in a industrial setting with positive feedback and we are leveraging constant developer feedback as a data flywheel to iteratively fine-tune and improve our models. 

Our contributions in this paper can be summarized as:
\begin{itemize}[leftmargin=20pt]
    \item We develop Code Quality Score (CQS), a system to automatically evaluate code quality and generate code reviews in a scalable manner.
    \item CQS is powered by open source LLMs (Llama models). We use a multi-stage post-training pipeline to better align with codebase context and developer preferences, and improve performance as compared to the base Llama models.
    \item To enhance end-to-end developer experience, CQS orchestrates multiple components for different stages of code review generation and validation, along with layers of sanity check before code reviews are send to developers.
    \item Overall, the CQS system shows an impressive precision rate for identifying valid issues with a set of code changes. It has been rolled out to more than 5000 engineers at Meta and has achieved a week-over-week user helpfulness rate of approximately $60\%$. 
\end{itemize}
Our hope is that systems like CQS can be used in a scalable way to evaluate codebase quality to alleviate some of the human review burden from senior engineers and as a feedback source for junior developers to learn about good coding practices in context of nuances of a specific codebase.

\section{Related Work}
\paragraph{AI Assisted Automated Code Review}
Initial attempts at automating aspects of the code reviewing process mostly leveraged pre-training recipes with transformer based model architectures. For example, \cite{tufano2022using} used source code and code reviews from GitHub to train an Text-to-Text Transformer (T5) model \citep{raffel2020exploring} while \cite{li2022automating} train an encoder-decoder model using code diff hunks (instead of using the source code directly) and corresponding reviews collected from high quality GitHub projects. However, more recently, the focus has shifted away from training models from scratch and towards leveraging world knowledge from foundation models such as Llama \citep{touvron2023llama} or Claude models, to create systems for code review. Unsurprisingly, using LLMs for code reviewing (and in general for coding tasks) has outperformed domain specific models trained only on code generation/code review tasks. For example, \cite{lu2023llama} design Llama-Reviewer, a framework aimed to fully automate the code review process by integrating code review generation and code refinement steps, by fine-tuning Llama models. 

Similarly, \cite{vijayvergiya2024ai} develop
AutoCommenter, an LLM-based system deployed at Google for detecting violations of coding best practices frequently referenced by human reviewers. Not only did AutoCommenter demonstrate the feasibility of deploying such an end-to-end system at scale, it also demonstrated high overall efficacy and user acceptance with over $50\%$ developer helpfulness rating. As compared to the Llama-Reviewer framework of \citep{lu2023llama}, we only restrict ourselves to generating high quality code reviews and go beyond training data collected from GitHub to fine-tune models using a multi-stage post-training pipeline comprised of SFT and offline RL approaches. As compared to AutoCommenter \citep{vijayvergiya2024ai}, we go beyond only detecting URL based best practice violations and instead focus on a variety of issues referenced by developers in their code reviews. In addition, CQS orchestrates multiple components to improve precision and developer experience, and is designed to leverage constant developer feedback to iteratively improve based on new issues that may arise in new code context (for e.g. when new frameworks are introduced in the codebase), developer preferences etc.


\cite{coderabbitai2024awesome} is another system similar to ours and offers AI-powered code review for enterprise customers by leveraging Anthropic's Claude series of models to provide context-aware feedback on pull requests with AST analysis and integration of static analyzers. While relying on external model providers introduces cost dependencies and limits customization, the CQS system is powered by open-source Llama models enabling controlled compute costs and deep customization.


\paragraph{Fine-tuning LLMs}
Large language models trained with self-supervised learning have shown impressive zero-shot/few-shot performance in a variety of tasks \citep{brown2020language, chowdhery2023palm}. With the explosion of LLMs and their applications in recent years, alignment/fine-tuning has emerged a major research area with the aim to improve performance of LLMs on specific tasks/human intent. There is a large body of work exploring novel methods for fine-tuning LLMs. For brevity, we only mention work that we directly leveraged. 

The Reinforcement Learning Fine-tuning (RFT) pipeline typically consists of an SFT phase, where a pre-trained LLM is fine-tuned with supervised learning on high quality data from the downstream task of interest, followed by RL optimization \citep{ziegler2019fine, ouyang2022training}. Both online RL (using reward models in PPO training) and offline RL approaches (by sampling multiple generations from the LLM and making preference pairs) exist for the RL fine-tuning phase. In our work, we only explore offline RL approaches based on Direct Preference Optimization (DPO) \cite{rafailov2023direct} due to its simplicity\footnote{We wanted to avoid training reward models with limited data we had.}. To create preference pairs for DPO, we used an LLM-judge model (that we also fine-tuned with SFT) to critique and score generations.

\paragraph{LLM-Judge for Preference Optimization}
As mentioned above, we use an LLM-judge model to create preference pairs for DPO training. Using LLM-judge models in the LLM development cycle has emerged as an important alternative to using human annotations (which can be costly and time consuming to obtain). Very recent work shown impressive results using LLM-judge models as reward models/scoring functions for creating preference pairs \citep{wang2024self, yuan2024self, kim2024aligning}.
Another closely related line of work has focused on judge models that first verbalize a critique and then map it to a preference score, yielding more transparent and often more data-efficient alignment signals. For example, the Critique-out-Loud approach of \cite{ankner2024critique} explicitly trains a judge model to verbalize free-form critiques, showing that such rationales can significantly improve pairwise preference classification accuracy.


We want to note two points in context of our work. First, for CQS, we were unable to use third party closed LLMs (like GPT 4, Claude models) as judge models to obtain feedback signal for training due to restricted use license and found off-the-shelf open models to under perform the LLM-judge model we trained using developer feedback (see section \ref{subsec:train_issue_validator} for details). Second, we did find the approach of using an LLM-judge model to be quite effective in the absence of human annotations -- our trained judge model was able to effectively transfer limited developer feedback into preference signals for fine-tuning.

\section{The Code Quality Score (CQS) system}
The code quality score (CQS) system orchestrates multiple LLM based components to automate the code review process -- with a slight abuse of the ``agent'' terminology, we refer to the CQS system architecture as being a ``multi-agent'' one. Such a multi-agent architecture allows distinct, specialized roles to be assigned to different agents. The CQS system architecture shown below in Figure  \ref{fig:system_design}, consists of two agents, namely, the `issue collector' agent and the `issue validator' agent. The issue collector agent uses an LLM to identify and summarize a set of possible issues with a given code change. The issue validator agent consists of an LLM that scores each issue collected by the issue collector agent, followed by a post-processing step which filters out some of the issues based on this score (using a threshold) as well as some rule based filters. In the following sections, we describe the prompt design and training of the LLMs used by the issue collector and issue validator agents.


\begin{figure}[h]
\label{fig:system_design}
\begin{center}
\includegraphics[width=\linewidth]{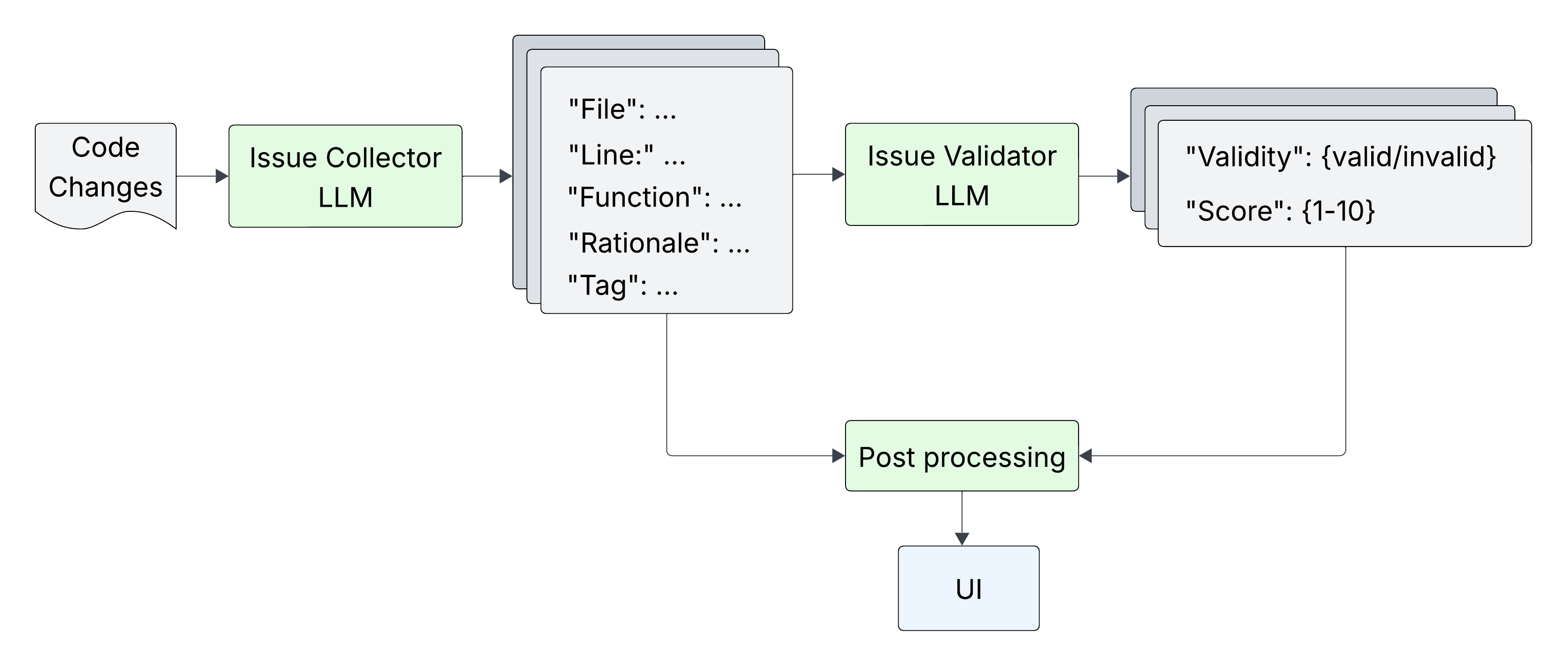}
\end{center}
\caption{Code Quality Score System Design}
\end{figure}

\subsection{A note on data format}
\begin{figure}[h]
\begin{center}
\includegraphics[width=1\linewidth]{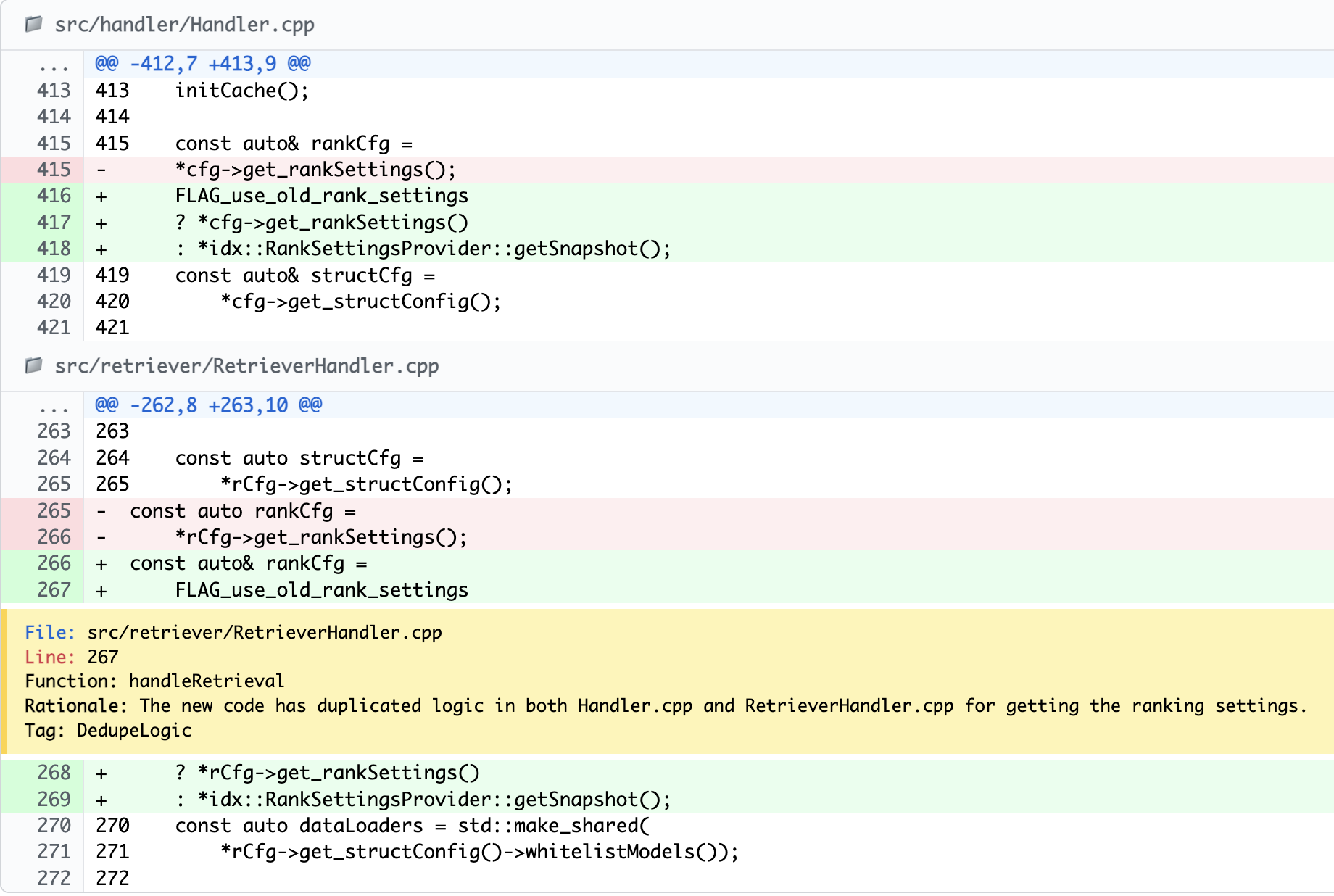}
\end{center}
\caption{How our system works: An example of code changes in diff format and code quality issues identified by LLM}
\label{fig:cqs_issue_ui_demo}
\end{figure}

We follow past work \citep{li2022automating, vijayvergiya2024ai} and use a ``diff'' format to represent code changes as inputs to the CQS system. An example of this is shown below in Snippet 1. The diff format is commonly used to show code changes clearly and is generated by comparing the original version and revised version of code. As compared to using the entire source code file, the diff format is an efficient representation as code changes are marked with ``+'' and ``-'' tags at the beginning of each line along with some surrounding line of unchanged code for context and file names. See \ref{fig:cqs_issue_ui_demo}

\subsection{The `issue collector' agent
}
The issue collector agent is based on a Llama 3.1-70b model. It scans the proposed code changes (diff format) and outputs a list of potential quality issues (for example, around code readability, complexity, robustness, modularity etc.). For this, we fine-tune a Llama 3.1-70b Instruct model using an initial training round of SFT followed by DPO training. We prompt the LLM to output a summary of the all possible issues identified by the model -- such a summary includes an issue tag, function name\footnote{For the function name, we prompt the model to output \texttt{NULL} in case the identified issue is with code change that is not a part of any function.}, rationale for the issue, file path and line number. We include a list of potential issue tags in the LLM prompt for the model to choose from but also take into account any novel tags that the model may occasionally come up with. See Appendix \ref{subsec:appendix_issue_collector_sys_instruct} for a detailed example of how we prompt the issue collector agent.



\subsection{The `issue validator' agent}
For a diff, the set of issues identified by the issue collector LLM are not directly shown in the UI. Instead, the outputs are sent to an issue validator agent, which uses another fine-tuned Llama 3.1-70b model under the hood, to critically access the validity and relevance of each identified issue, and outputs a score on a scale of 0-10. For some issues (based on issue tag), we include specific guidelines in the prompt to help the model focus on particular elements when validating these issues.
The outputs of the collector and validator agents are post-processed to filter out issues using a combination of rule based filtering logic and the score outputted by the validator agent, before the code review is shown in the UI. Next, we give details of how the collector and validator LLMs were fine-tuned.

\subsection{Model Training}
We describe in detail the training paradigm we use to fine-tune the issue collector and validator LLMs. As mentioned before, both these LLMs are fine-tuned using the Llama 3.1 70b Instruct model as the base checkpoint.


\subsubsection{Issue Collector LLM}
\label{subsec:train_issue_collector}
The issue collector LLM is trained in two phases -- the initial model is fine-tuned using SFT -- which is followed by a round of offline-RL based fine-tuning using DPO. This is a commonly used recipe for fine-tuning foundation models -- our motivation for this choice was to teach the model about developer preferences using SFT and then do DPO training for better generalization\footnote{The preference data for DPO training was sampled from the model itself. By doing a round of SFT before, our goal was to be able to capture developer preferences in the DPO training data as well.}. We also do ablations with the SFT model and a DPO model trained without doing SFT first. Doing a round of SFT training before DPO seems to help improve the recall rate. See Section \ref{subsec:issue_collector_llm_finetuning} for details. 

\paragraph{Phase 1: Supervised Fine-tuning (SFT)}
For SFT training, we have collected large amount of human reviewed "diffs" from Meta’s internal codebase in recent 2 years, the diffs covers a wide range of tasks/projects/services. We have applied a variety of rules to filter down to a dataset of code changes which were identified to contain good quality human reviews. The set of code changes we collect cover a wide ranging set of tasks, projects, and services, including 7 different programming languages, with the majority of code changes evenly split between Python, C++ and PHP. For dataset creation, we do apply some filters -- for example, we only consider diffs with more than 1 review comment and discard diffs authored by a bot or those that were solely created for testing purposes. In addition, we only selected code changes that were aimed specifically to solve code quality issues\footnote{The kind of issues around `Readability and Maintainability', `Modularity and Reusability' etc. that we want to identify. For reference, please see the prompt we use for the issue collector LLM in Appendix \ref{subsec:appendix_issue_collector_sys_instruct} for details about the major issue categories that we aim for the issue collector LLM to identify.} and curated a total of approximately 6000 diffs. For each diff, we classify every human reviewer comment into ``tags'' of possible issue the human reviewer may be referring to and create a ``code review'' which contains issue tag, comment by the human reviewer along with the function name (\texttt{NULL} if the comment does not refer to any function), file path and line number for each issue. 

However, even in this curated dataset of (diff, code review), we found a considerable portion of examples where at some lines of the code change, the human review comments were not useful or did not have any clear guidance. To do further filtering, we use a Llama 3.1 70b Instruct model by prompting it to analyze the quality of the human reviewer comments and to grade them (classify as ``good'' or ``bad''). We also instruct the model to rewrite the human comments into ``rationales'' for the issue. See Appendix \ref{subsec:appendix_rewrite_human_comments_sys_instruct} for details about the prompt we use. From each code change, we dropped issues where human comments were classified as ``bad'' by the Llama model and obtained a set of 5000 (diff, code reviews), each with details about high quality issues derived from human comments. We use this dataset to do supervised fine-tuning (SFT). The entire data curation and training pipeline for Phase 1 is illustrated below in Figure \ref{fig:issue_collector_llm_phase_1_training}.

\begin{figure}[h]
\begin{center}
\includegraphics[width=1.1\linewidth]{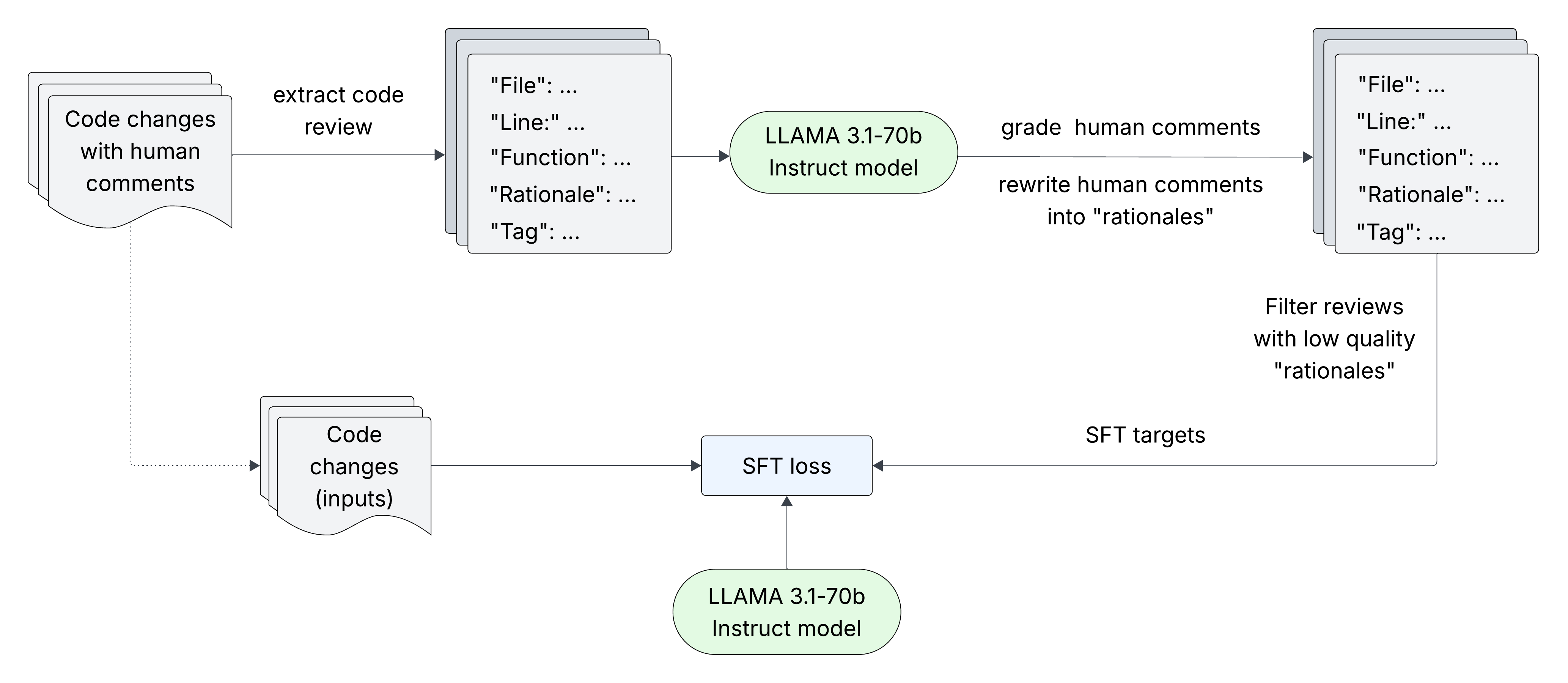}
\end{center}
\caption{Phase 1 training of the Issue Collector LLM using Supervised Fine-tuning}
\label{fig:issue_collector_llm_phase_1_training}
\end{figure} 

\paragraph{Phase 2: Offline RL using Direct Preference Optimization (DPO)} 
We further fine-tune the SFT checkpoint (obtained after phase 1 training) using DPO \citep{rafailov2023direct}, an offline RL algorithm popularly used for reinforcement learning fine-tuning. DPO training requires data in the form of preference pairs, consisting of a `winner' and a `loser' response for each input, and attempts to increase the log probabilities of tokens in the `winner' response while decreasing the log probabilities of tokens in the `loser' response. Concretely, for a dataset $\mathcal{D}$ of preference pairs, the training objective for DPO is given by:

\begin{equation*}
\label{eq:dpo_loss}
\mathcal{L}_{\text{DPO}}(\pi_\theta; \pi_{\text{ref}}) = -\mathbb{E}_{(x,y_w,y_l) \sim \mathcal{D}} \left[ \log \sigma\left(\beta \log \frac{\pi_\theta(y_w|x)}{\pi_{\text{ref}}(y_w|x)} - \beta \log \frac{\pi_\theta(y_l|x)}{\pi_{\text{ref}}(y_l|x)}\right) \right],
\end{equation*}

where $\pi_\theta$ is the policy being trained, $\pi_{\text{ref}}$ is the reference model (in our case it is the SFT policy obtained after Phase 1 training), $x$ represents the input (in our case, the code changes in form of a diff), $y_w$ and $y_l$ are the `winner' and `loser' responses respectively (in our case, a pair of correct and incorrect issues identified in the code review), and $\beta$ is the temperature parameter controlling the strength of the KL penalty.

To create preference pairs, we used an LLM-judge to score generations from the SFT model (essentially, to score each issue identified in the model generated code review). An alternate way to score generations is to use human annotations instead -- however, obtaining such human feedback was less scalable for us (we found developers less willing to compare and judge two code reviews) and so we relied on the LLM-judge model. To sample diverse generations, we rely on using high temperature sampling only. Recent work in literature \citep{nguyen2024turning, hsu2025group} has explored alternate methods to improve diversity of the sampled generations which could be interesting to explore in our setting as well -- we leave this exploration for future work.


We prompt the LLM-judge to score each issue in a code review (i.e. each issue in $y_i^{(j)}$ in Equation \ref{eq:judge_scoring} below) on a scale of 0 to 10 -- a score of 0 if the issue identified was found to be entirely incorrect or empty, and a score between 1 to 10 to grade the quality/correctness of the issue identified. See Appendix \ref{subsec:appendix_llm-judge_scoring_sys_instruct} for the detailed prompt we use to score generations using LLM-judge.

\begin{equation*}
\{y_i^{(1)}, y_i^{(2)}, \ldots, y_i^{(N)}\} \sim \pi_{\text{SFT}}(\cdot | x_i)
\label{eq:response_sampling}
\end{equation*}
\begin{equation}
s_i^{(j)} = \text{LLM-Judge}(x_i, y_i^{(j)}) \in [0, 10]
\label{eq:judge_scoring}
\end{equation}

The pseudo-code for creating preference pairs using LLM-judge scores is shown in Algorithm~\ref{alg:gen_pref_pairs} below. For each code change, we compare all pairs of model generated code reviews and for each pair, we compare issues with the same issue tag but different scores (as graded by the LLM-judge) -- if the score difference is greater than a threshold (chosen empirically), we add this pair of issues to our training dataset for DPO training. This way, we were able to obtain a set of 8400 preference pairs using the set of 5000 diffs we had curated (we used the same set of diffs as we used for SFT training in phase 1). The data creation pipeline we used for DPO is shown below in Figure \ref{fig:train_issue_collector_llm_phase_2}.

\begin{algorithm}
\caption{Make Issue-wise Preference Pairs for DPO}
\label{alg:gen_pref_pairs}
\begin{algorithmic}[1]
    \State \textbf{Inputs}: code changes $x_i$, model generated code reviews $\{y_i^{(j)}\}_{j=1}^{10}$, threshold $\delta$, preference\_pairs $\gets$ [ ]
    \For{each $x_i$}
    \State $\text{issue\_list}(y_i^{(j)})$ $\gets$ extract list of issues from each code review $y_i^{(j)}$.
        \ForAll{$\{(a, b) \,\, | \,\, a \in \text{issue\_list}(y_i^{j}), b \in \text{issue\_list}(y_i^{k})$\}}
            \If{$a.\text{tag} = b.\text{tag}$ \textbf{and} $|a.\text{score} - b.\text{score}| \geq \delta$}
                \State preference\_pairs.append(\{$x_i$, chosen=$a$, rejected=$b$\})
            \EndIf
        \EndFor
    \EndFor
    \State \Return preference\_pairs
\end{algorithmic}
\end{algorithm}


\begin{figure}[h]
\begin{center}
\includegraphics[width=1.1\linewidth]{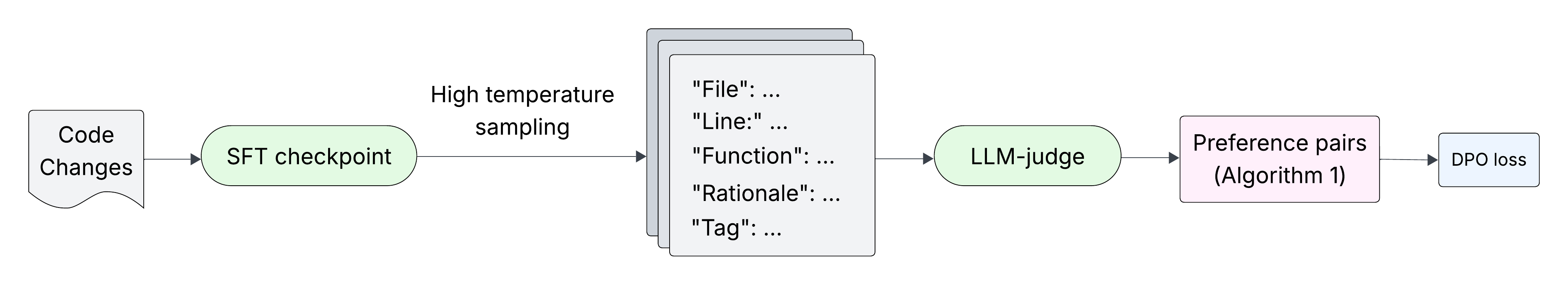}
\end{center}
\caption{Phase 2 training of the Issue Collector LLM with DPO using LLM-judge to create preference pairs}
\label{fig:train_issue_collector_llm_phase_2}
\end{figure}

We describe training of the LLM-judge below in Section \ref{subsec:train_issue_validator} since we also use the LLM-judge model as a part of the validator agent.

\subsubsection{Issue Validator Agent}
\label{subsec:train_issue_validator}
The issue validator agent consists of the LLM-judge model and a post-processing step to filter out incorrect issues before code reviews are shown in the UI.

\paragraph{LLM-judge model:} Our motivation for training an LLM-judge model and use it for creating preference pairs for DPO training was that we we found developers more willing to "critique" individual responses rather than eyeball and compare two responses. Therefore, we aimed to distill this developer feedback into an LLM-judge model.
To do this, we follow a recent proposal by \citet{ke2023critiquellm} and fine-tune a Llama 3.1-70b Instruct model to generate good ``critiques" of a code review, to imitate how a human developer would judge a code review. Recall that we use the LLM-judge model to score code review responses sampled from the SFT model on a scale of 0-10; therefore, we need the LLM-judge model to provide fine-grained distinguish-ability between different sampled responses. The key finding of \citet{ke2023critiquellm} is that state-of-the-art LLMs like ChatGPT, GPT-4 etc. lack the ability to generate informative ``critiques" to evaluate and grade LLM-generated responses on different real-world NLP tasks such as AlignBench \citep{liu2023alignbench} and LLMEval \cite{zhang2024llmeval}. However, fine-tuning on a small data set of good quality ``critiques'' improves a models ability to generate granular scores on LLM-generated responses. The resulting model tends to not only perform well on evaluation tasks but can also be used to provide scalable feedback in an fine-tuning loop. 

The data curation and training pipeline for the LLM-judge model is illustrated below in Figure \ref{fig:llm_judge_training}. To acquire ``critiques'' for LLM-judge training, we rolled out the SFT model (obtained after phase 1 training) to a group of developers for beta testing. For each code change, the SFT model generated a code review with different identified issues, and developers were asked to provide both textual feedback (``critique'' of each issue identified by the model) and click a thumbs up/down button (to indicate if they agreed or disagreed with the issue identified). This way we were able to collect developer feedback in the form of (``critiques'', ``grades'') pairs for every issue for a set of approximately 1500 diffs. 

However, we observed that not all developer critiques were of high quality and thus we used another Llama 3.1-70 Instruct model to analyze the critiques written by developer, rewrite them and score (see Appendix \ref{subsec:appendix_human_feedback_critques} for the exact prompt we use). We then filter away developer feedback which was judged to have low quality critiques and use the remaining developer feedback for LLM-judge model training using SFT. For this round of SFT, we use code changes and the corresponding model generated code reviews as inputs, and the curated set of code review critiques obtained from developer feedback as SFT targets. 


\begin{figure}[h]
\begin{center}
\includegraphics[width=1.1\linewidth]{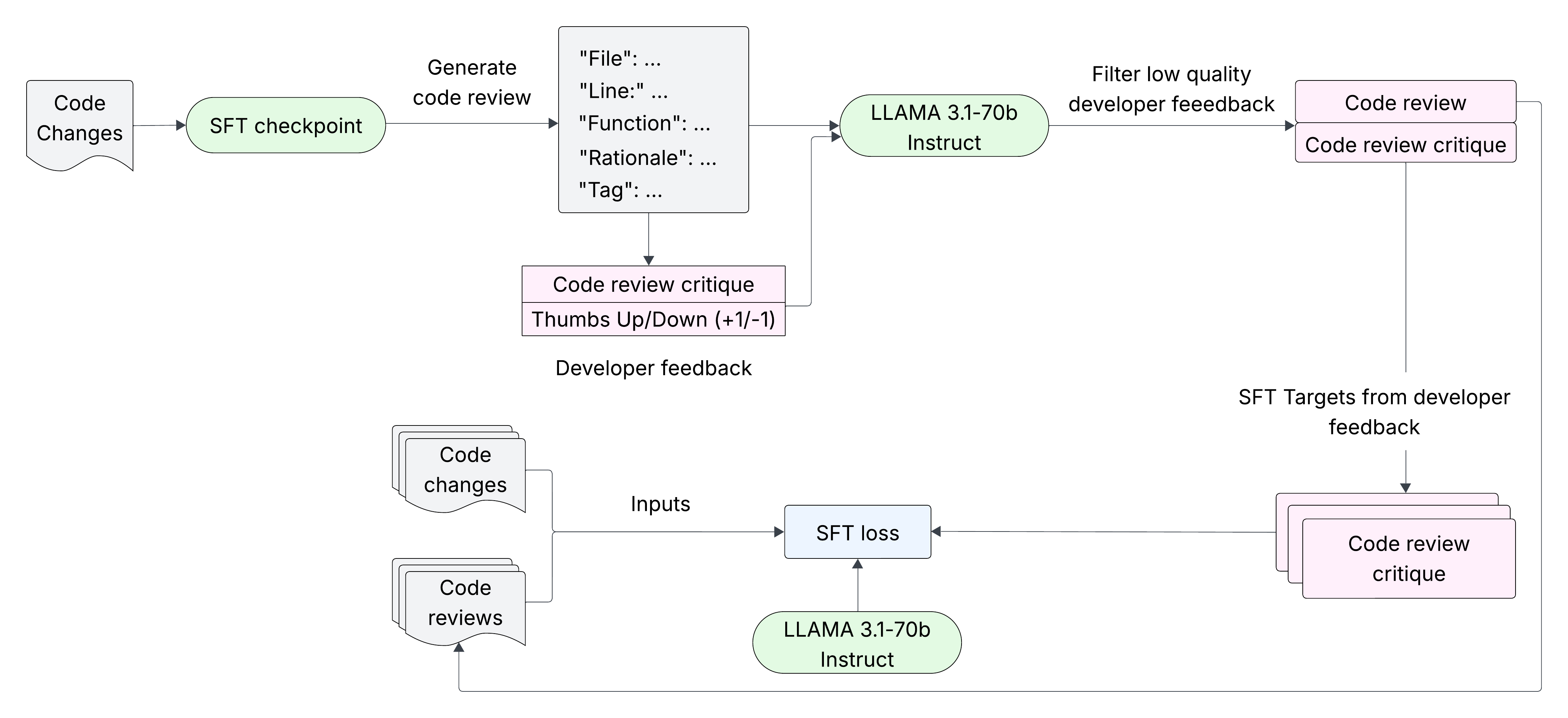}
\end{center}
\caption{LLM-judge training. We curate high quality developer feedback (code review critique, thumbs up/down) for model generated code reviews sampled from the SFT checkpoint of the issue collector LLM. This developer feedback is then used to SFT a Llama model to be good at generating human like critiques and grades for model generated code reviews.}
\label{fig:llm_judge_training}
\end{figure}

\paragraph{Post-processing:} 
Even though the LLM-judge we trained is able to evaluate LLM-generate code reviews well (significantly better than the base Llama 3.1-70b Instruct model), it isn't perfect. Therefore, the issue collector LLM, trained on a limited dataset of human code reviews and preference pairs created by LLM-judge scores, can sometimes generate code reviews which are inaccurate and misleading 


To counter this, we add layers of sanity checks before sharing the code review in the UI. For example, we score all code reviews generated by the issue collector LLM using the LLM-judge model (see Appendix \ref{subsec:appendix_issue_validator_sys_instruct} for the exact prompt we use) and take this score into account as a filter to check if the code review passes a quality threshold. In addition, we also have numerous hard coded filters specific to each issue tag and programming language as we observed empirically the issue accuracy varies across tags and languages. The code review needs to pass all these filters to be shown in our internal UI and shared with the author.

\section{Evaluation Design}
\label{sec:evaluation}
To create an evaluation dataset for CQS, we designed a pipeline which uses multiple frontier LLMs\footnote{We use multiple LLMs to maximize coverage of the possible issues that can be identified.} such as GPT-4o, Claude 3.7 Sonnet and Gemini-2.5-pro to generate code reviews for a small set of code changes (separate from those used as part of training data). We then use human annotators to analyze and label each issue identified in the code reviews, and produce a small set of (diff, code review) pairs which we use as the evaluation benchmark. During evaluation runs, we use semi-deterministic verification to produce evaluation metrics.

 
\vspace{-0.5em}
\paragraph{LLM based code reviews:}
To generate code reviews from state-of-the-art LLMs, we collect a set of 200 diffs that cover a diverse set of projects, across 7 programming languages with a majority of these being either C++, PHP or Python. As mentioned above, to maximize issue coverage we use GPT-4o, Claude 3.7 Sonnet and Gemini-2.5-pro to generate ``candidate'' code reviews.
\vspace{-0.5em}
\paragraph{Human annotation and evaluation dataset:} After generating code reviews using state-of-the-art LLMs, we engage with multiple engineers to review the issues identified. Each code change is assigned to at least two engineers. For each code review, we only keep issue where both engineers agree on issue tag, line number and issue rationale -- issues on which engineers disagree are discarded which ensures trustworthiness of our evaluation dataset. This resulted in a final set of approximately 80 (diff, code review) pairs with each containing approximately 2-4 ``consensus'' issues identified.
 
\vspace{-0.5em}
\paragraph{Evaluation protocol and metrics:}
During evaluation runs, we apply both rule based and semantic matching to compare code reviews generated by the CQS system with those in the evaluation dataset, issue by issue. We do this by constructing an automatic verifier that first performs an exact string matching check for issue tags. If the tag match is successful, we compare the rationale for each issue -- this is done using a lightweight Llama 3.1-8b model given the comparison task here is simple\footnote{We didn't use BLEU score as done in \cite{li2022automating}, given the cost of using a lightweight Llama 3.1-8b model during evaluation runs is small.}. In addition, we also eyeballed some issue comparisons as a sanity check for our metrics. 

\paragraph{Bias in evaluation data:}
A challenge in evaluating quality of code reviews  generated from a system like CQS is that ground truth can be incomplete and biased, since human annotations can be insufficient and often noisy (\cite{vijayvergiya2024ai};~\cite{li2022automating}). We also struggle with similar issues. Since we use GPT-4o, Claude 3.7 Sonnet and Gemini-2.5-pro to collect issues, the evaluation data is heavily biased towards these models. While we asked multiple senior engineers to validate each issue, this did not get rid of the bias since we did not ask them to identify possible issues which might have been missed by the frontier LLMs. In addition, during validation, engineers were also able to see the model name for each identified issue -- we think this might have further amplified bias towards the Claude 3.7 Sonnet model.

\section{Results}
In this section, we present evaluation results. We primarily evaluate the CQS system on two metrics -- \textit{precision}, i.e the fraction of issues identified by CQS that were also identified in the evaluation dataset that we curated, and \textit{recall}, i.e. the fraction of issues identified in the evaluation dataset that CQS is also able to successfully identify. The CQS system design and prompt design is aimed at improving precision while fine-tuning different components is aimed at optimizing for both precision and recall. We do note that these results should be read in context of the bias in the evaluation dataset as mentioned above in Section \ref{sec:evaluation}. Despite this, the CQS system shows impressive precision rate as shown in Table \ref{tab:precision_growth}.

\subsection{Fine-tuning the issue collector llm}
\label{subsec:issue_collector_llm_finetuning}
In Table \ref{tab:finetuning_results} below, we compare the precision and recall rates for different fine-tuning recipes we tried while training the issue collector LLM. Note, we report the numbers averaged over 10 evaluation runs. As can be seen, the SFT trained model seems to be at par (even slight regression in terms of recall rates) as compared to the base Llama 3.1-70b Instruct model. However, DPO trained models do show improvements. While directly fine-tuning the base model with DPO shows a 3\% lift in precision, it does not improve recall. On the other hand, a combination of SFT and DPO improves both precision  and recall rates by 2\% and 1\% respectively indicating improved generalization. A takeaway from our results is that using an LLM-judge model (even with very limited data) seems to have worked -- it will be interesting to see the performance gains that can be achieved by improving the LLM-judge model using more developer feedback. We leave this investigation for future work. 

A noteworthy observation from our results is that fine-tuning using DPO directly from the base model yields a lower recall rate than following the SFT + DPO recipe. A possible explanation could be that doing SFT on human code reviews as a first step likely helps the model mimic developer preferences more closely. Another interesting observation is that only doing DPO training results in the model identifying smaller number of issues overall as compared to the base Llama model as well as the SFT model. Recall that during DPO training, we created ``winner'' and ``loser'' samples in each preference pair with exactly one issue -- so the training data does not have any bias towards smaller number of issues. We found this to be a bit surprising but do not have a very good explanation for it. 

The huge gap between Llama 3.1 70b models and other state-of-the-art models like Claude 3.7 Sonnet, GPT-4o and Gemini 2.5-pro is also noteworthy. Although the Gemini 2.5-pro model has a better precision rate, Claude 3.7 Sonnet seems to be the best performing model overall. We would like to remind the readers again of the caveat that since these models were used as part of the evaluation dataset creation (see Section \ref{sec:evaluation}), it is very likely that the precision and recall rates are biased. Nevertheless, the numbers do show a significant advantage of the Claude, GPT and Gemini models.



\begin{table}[h]
\centering

\begin{tabular}{|c|c|c|c|}
\hline
\textbf{Model} & \textbf{\# issues found} & \textbf{precision (\%)} & \textbf{recall (\%)} \rule{0pt}{10pt} \\
\hline
Baseline Llama-3.1-70b Instruct & 138 & 11.28 & 8.14 \rule{0pt}{10pt} \\
\hline
Llama 3.1-70b Instruct + SFT & 130 & 11.14 & 7.62 \rule{0pt}{10pt} \\
\hline
Llama-3.1-70B Instruct + DPO & 103 & 14.56 & 7.82 \rule{0pt}{10pt} \\
\hline
Llama 3.1-70B Instruct + SFT + DPO & 131 & 13.48 & 9.25 \rule{0pt}{10pt} \\
\hline
Claude 3.7 Sonnet & 193 & 30.43 & 30.58 \rule{0pt}{10pt} \\
\hline
GPT-4o & 233 & 24.33 & 29.53 \rule{0pt}{10pt} \\
\hline
Gemini-2.5-pro & 134 & 32.77 & 22.53 \rule{0pt}{10pt} \\
\hline
\end{tabular}
\caption{Evaluation metrics: precision \& recall rates across different stage of training. Here ``\# issue found'' refers to the total number of issues identified by the model over the entire evaluation dataset.}
\label{tab:finetuning_results}
\end{table}

\subsection{Multi-stage Prompt Optimization}
While designing the CQS system, we prioritize precision over recall to ensure developer trust in the service\footnote{Developers want the system identified issues to be valid and are fine with the system not being able to identify all issues.}. In Table \ref{tab:precision_growth} below, we show how the precision rates change for different system configurations. As compared to only using the issue collector agent, using the issue validator agent (LLM-judge scores along with rule based filters) is really helpful in improving the precision rate of the CQS system. 

We also benchmark against a setting where we use two Claude 3.7 Sonnet models, one for issue collection and the other as LLM-judge for scoring the collected issue, along with post-processing using rule based filters. As can be seen in Table \ref{tab:precision_growth} below, while this does lift precision (from 30.43\% to 51.63\%), it lags in comparison to the CQS system using fine-tuned Llama models. The LLM-judge model used for scoring (as part of issue validator agent) seems to be responsible for this gap -- off-the- shelf Claude 3.7 Sonnet model is unable to distinguish between valid and invalid issues as compared to the Llama 3.1 70b LLM-judge model we trained. This is expected as prior research shows fine-tuning large language models (even with a small dataset of high quality) can be effective in improving its accuracy as a judge model \citep{ke2023critiquellm, wang2024self}.


\begin{table}[h]
\centering
\begin{tabular}{|C{7cm}|C{3cm}|C{2.5cm}|}
\hline
\rowcolor{lightgray}
\textbf{Method} & \textbf{Avg precision (\%)} & \textbf{Avg recall (\%)} \rule{0pt}{12pt} \\
\hline
Issue collector agent & \makecell{13.48} & \makecell{9.25} \rule{0pt}{12pt} \\
\hline
Issue collector and Issue validator agent (including post-processing filtering) & \makecell{78.20} & \makecell{1.20} \\
\hline
Claude 3.7 Sonnet (for both issue collection and issue validation) and post-processing filtering & \makecell{51.63} & \makecell{11.45} \rule{0pt}{12pt}  \\
\hline
\end{tabular}
\caption{Precision/recall rates through the different stages of the CQS system. 
}
\label{tab:precision_growth}
\end{table}

\subsection{llm-judge evaluation}
Recall that we trained the LLM-judge model using developer feedback on code reviews generated by the SFT checkpoint. From this developer feedback, we also create a small dataset of 200 code reviews with thumbs up/down signal to evaluate the LLM-judge model -- with 100 positive examples (code reviews with a thumbs up) and 100 negative examples (code reviews with a thumbs down). On this evaluation data set, our LLM-judge model scored a 62\% accuracy as compared to a 56\% accuracy scored by a Llama 3.1-70b Instruct model. We give details of a small ablation study on the effect of SFT for LLM-judge model training in 
Appendix~\ref{sec: judge_study}.

\subsection{Online Results}
CQS has been rolled out to a large number of developers with consistently good user feedback. For instance, it has an average 60\% WoW user helpfulness rate (WoW helpfulness rate is the percentage of positive feedback that developers have shared by clicking a thumbs-up/down button in the code review UI). While we see a correlation between improvements in precision and user helpfulness rate, it is not directly proportional. A potential reason could be the small size of our evaluation dataset -- model selection based on improvements in our evaluation metrics may not transfer when the system is rolled out to a diverse set of code changes. We currently observe a $\pm 10\%$ confidence interval on the WoW user helpfulness rate and are working on increasing the size, quality and coverage of our evaluation dataset.

\subsection{Discussion}
We note a couple of observations from our experience designing the CQS system. First, we found LLM-judge model performance to be critical for DPO training. This is expected -- reward/feedback quality is known to be the key component in RL based fine-tuning of LLMs and more generally in RL as well. Training high quality reward models with limited feedback data (such as our context) seems to be an important research question. While current work has leveraged textual reasoning using critiques \cite{ankner2024critique} or supervised fine-tuning techniques \cite{ke2023critiquellm, wang2024self} to enhance performance of LLM-judge models, other ideas such as using reasoning models with multiple evaluation rubrics might be useful in training robust reward models. 

Second, the idea of using an LLM-judge model and rule based filtering as part of the issue validator agent was motivated by developer expectations of high precision rates from a system like CQS. As can be seen in Table \ref{tab:finetuning_results}, even frontier models have low stand-alone precision rates ($<33\%$) for identifying code quality issues. It will be interesting to see how these precision rates improve with the next generation of frontier models. Despite improvements we made with the CQS system, extracting high-quality, actionable code review comments for diff remains a significant challenge.

\section{Conclusion and Future Work}
The Code Quality Score (CQS) system we designed has shown good performance with early adoption by large number of developers and good feedback with a 60\% week-over-week user helpfulness rate. Although Llama models lag state-of-the-art closed source models like Claude model series on issue collection tasks, we were able to fine-tune them using developer feedback to improve their performance on issue collection and validation tasks, and leverage them effectively in a multi-agent system to achieve a better precision rate overall (with issue validator agent). As Llama models improve and close the gap with other state-of-the art models, we hope to further improve the performance of our system. 

We also plan to further enhance the CQS system by leveraging online developer feedback to build a data flywheel that can be used to continuously improve Llama models on issue collection and validations tasks. Another direction for future work that we plan to pursue is to go beyond code quality issues and expand the CQS system to identify complex issues related to security vulnerabilities and performance.

Although AI driven code review systems are still in their early stages, early results are encouraging in terms of adoption and satisfaction rates. As these systems mature, they are not only expected to reduce human review burden but also bring long-term improvements to the codebase health -- although quantifying these benefits is a complex exercise that requires continuous data collection and monitoring. We leave such studies for future work.

\section{Acknowledgement}
We thank the team members and leadership in Meta's monetization organization for their support in developing and deploying the Code Quality Score system, including Wenlin Chen, Hung Duong, Shah Rahman, Ritwik Tewari, Santanu Kolay, Neeraj Bhatia, Matt Steiner, Peng Fan

\newpage
\bibliography{iclr2025_conference}
\bibliographystyle{iclr2025_conference}

\newpage
\appendix
\section{Appendix: different prompts used}
\label{sec:appendix_sec_prompts}
We list the prompts that were used for different LLMs.

\subsection{System instruction for the issue collector llm}
\label{subsec:appendix_issue_collector_sys_instruct}
\begin{tcolorbox}[colback=white,  
                   colframe=black,    
                   coltext=blue,      
                   breakable,
                   fontupper=\fontfamily{pcr}\selectfont]  

=== Raw Source Control Code Changes BEGIN === \\
code changes go here \\
=== Raw Source Control Code Changes END ===

\vspace{1em}
=== Context for the Code Change BEGIN === \\
The following additional information from the author may provide context about the code changes and their purpose: \\
Title: title \\
Summary: summary of code changes \\
=== Context for the Code Change END ===

\vspace{1em}
Your task is to provide constructive and concise feedback for the code changes based on the following criteria (each goes with a tag name in the beginning):
\begin{itemize}[leftmargin=15pt]
\item DedupeLogic: Deduplicate logic into shared functions except for logging.
\item DictionaryKeyExistenceCheck: Check for dictionary key existence before accessing it.
\item UseConstant: Use constants instead of literals, unless it's a constant definition.
\item RenamingVariable: Variable names should be pronounceable, easily readable, and reveal intent.
\item DomainSpecificName: Use solution domain names, computer science (CS) terms, algorithm names, pattern names, math terms. When there is no name from the solution domain then prefer problem domain names.
\item BreakdownLongFunction: Break down long functions into smaller, more focused functions. Only include this issue if the length of the current function is longer than 50 lines.
\item ExtractMethod: When we have a block of code that can be extracted into a separate method, we should do so.
\item ResourceLeak: A resource handle (e.g., file, socket, connection) is not properly wrapped in a context manager or try-with-resources construct or use a with statement in Python.
\item Documentation: Docstring should match the code, and be added for each major unit. Pay extra attention to different types of docstrings.
\item DivisionByZero: When performing division operations, ensure that the divisor is checked to be non-zero before proceeding with the calculation.
\item Typo: A typo is detected in the code.
\item RenamingFunction: Function names should be pronounceable, easily readable, and reveal intent.
\end{itemize}
\vspace{1em}
In case none of the provided tags are appropriate, you may come up with your own tag.
\vspace{1em}

Code lines are prefixed with the line number, followed by symbols ('+', '-', ' '). The '+' symbol indicates new code added, the '-' symbol indicates code removed, and the ' ' symbol indicates unchanged code. The review should address new code added (lines starting with '+'). \\
\vspace{1em}

Pay attention to what part of the code has changed, you should only grade the code that has changed, and ignore the rest.

\vspace{1em}
Example output:
\vspace{1em}
\begin{verbatim}
{
  "function": "[exact function name from code]",
  "rationale": "[clear explanation of the issue and suggested 
  improvement]",
  "file": "[file path where the function is located]",
  "line": [numeric line number],
  "tag": "[one of the tags as described above]"
}
\end{verbatim}

\vspace{1em}
Note you should not be overly critic and generate many bullet points. Please pin-point the `problematic\_function' with issues, and don't forget to include line numbers. The output field `problematic\_function' is mandatory in the response if the problem is inside a function. The output field `relevant\_file' should include full path of the file (including directory and filename). In your response, do not describe what the code changes is about because the code author already knows about it. Pay attention to the code change. \\

\vspace{1em}
The output field `rationale' would be shown to the author of code changes. Please use a polite and suggestive tone for the `rationale' field, provide a reasoning and give suggestions using a question instead of a command.
\end{tcolorbox}

\newpage
\subsection{System Instruction for Rewriting human comments into ``issue rationales"}
\label{subsec:appendix_rewrite_human_comments_sys_instruct}
\begin{tcolorbox}[colback=white,  
                   colframe=black,    
                   coltext=blue,      
                   breakable,
                   fontupper=\fontfamily{pcr}\selectfont]  

You are given a code change and a corresponding code review. 

\vspace{1em}
Here are the code changes: \\
=== Raw Source Control Code Change BEGIN === \\
code changes go here \\
=== Raw Source Control Code Change END ===

\vspace{1em}
Here is the code review in yaml format: \\
=== Code Review (YAML FORMAT) BEGIN === 
\begin{verbatim}
```yaml
{
  "Function": ...,
  "Rationale": ...,
  "File": ...,
  "Line": ...,
  "Tag": ...
}
\end{verbatim}
=== Code Review (YAML FORMAT) END ===

\vspace{1em}
Please  reason and understand the code review and carefully follow the instructions below.

\vspace{1em}
First, please do an initial assessment of the code review to try to roughly understand what the reviewer's objective. Then, try to explain if the rationale in reviewer's comments is logical. You may or may not have a clear understanding of what the rationale might be, so you can propose multiple hypothesis. For example: \\ \\
``Hypothesis: In file xyz.py, at line 100, the reviewer is referring to an issue in the definition of  `ProblematicFunction', and if the reviewer's comment is not addressed, it could cause X/Y/Z''

\vspace{1em}
After you propose hypothesis, reflect on them and answer the following questions:
\begin{enumerate}[leftmargin=20pt]
    \item Is the provided rationale addressing the issue in the give code change or it's referring to some context outside of the given code change?
    \item Is the rationale linking to external references/links?
    \item Is the rationale explicitly pointing out issues? Or is it just an inquiry because the review is not sure either?
    \item Is the reviewer's suggestion actionable? Or is it just some general suggestions that does not have clear action items?
    \item Is the reviewer nitpicking? For example, saying something like "nit...."
\end{enumerate}

\vspace{1em}
After you have answered these questions, you should have a clear judgment of the "quality" of the review. Typically, bad quality reviews have the following characteristics:
\begin{itemize}[leftmargin=20pt]
    \item Out of context: reviews which link to external references or refer to something outside of the given code change.
    \item Nitpicking.
    \item Not actionable.
    \item Reviews which do not detect SPECIFIC issues but are more general suggestions or free discussions.
\end{itemize} 

\vspace{1em}
If the review quality is good, please rewrite the rationale from an AI Assistant's perspective. Here are the hints to derive a good rationale:
\begin{itemize}[leftmargin=20pt]
    \item Try to VERIFY each hypothesis you proposed, using facts in the code change content.
    \item Summarize the verified hypothesis
    \item Try to refine the summary to a concise form, the final derived rationale should be specific, helpful and actionable.
\end{itemize}

\vspace{1em}
Finally, please output a <conclusion> in the following format. Note, if it's a bad quality review, you can set `review\_rewrite` field as null.

\begin{verbatim}
{
<conclusion>
   review_quality: {good/bad}
   review_rewrite: {human comment rewritten by an AI assistant}
</conclusion>
}
\end{verbatim}

\vspace{1em}
Please stop after writing the <conclusion> and don't output anything else.
\end{tcolorbox}

\newpage
\subsection{Prompt to rewrite human feedback into critiques for llm-judge training}
\label{subsec:appendix_human_feedback_critques}

\begin{tcolorbox}[colback=white,  
                   colframe=black,    
                   coltext=blue,      
                   breakable,
                   fontupper=\fontfamily{pcr}\selectfont]  
You are given a code review and a corresponding feedback to the review from developers. Your task is to analyze the developer feedback and grade it, in context of the code changes and code review.

\vspace{1em}
Here are the code changes: \\
=== Raw Source Control Diff BEGIN === \\
code changes go here \\
=== Raw Source Control Diff END ===

\vspace{1em}
Here are the code review comments for code changes above
\begin{verbatim}
{
  "Function": ...,
  "Rationale": ...,
  "File": ...,
  "Line": ...,
  "Tag": ...
}
\end{verbatim}

\vspace{1em}
And here is the human feedback to the code review. 
\begin{verbatim}
{
  "human sentiment": {positive/negative (thumbs up/down)}
  "human feedback comments": {}
}
\end{verbatim}
Note that "human feedback comments" could be empty if the user did not provide any. The "human sentiment" is positive if they agree with the review and negative otherwise.

\vspace{1em}
Please make sure you understand the code change, code review and the human feedback. Then, come up the following:
\begin{enumerate}[leftmargin=20pt]
    \item human\_feedback\_quality: a score from 0-5 about the quality of human feedback. A good feedback should specifically point out the problem in the code review. Output a score of 0 if you did not understand the feedback at all.
    \item critique: rewrite the human feedback to one sentence critique by using your knowledge of software engineering. If the feedback is of low quality or not available, please write your own based on the positive or negative sentiment.
\end{enumerate}

\vspace{1em}
Write your thoughts, and then respond with the following yaml format:
\begin{verbatim}
```yaml
"human_feedback_quality": {0-5, 0 means you did not understand  
    the human feedback at all}
"critique": {one sentence critique to the code review}
```
\end{verbatim}
\end{tcolorbox}

\newpage
\subsection{System Instruction for llm-judge scoring}
\label{subsec:appendix_llm-judge_scoring_sys_instruct}
\begin{tcolorbox}[colback=white,  
                   colframe=black,    
                   coltext=blue,      
                   breakable,
                   fontupper=\fontfamily{pcr}\selectfont]  
You are a language model that specializes in evaluating reviews for a code change. You are given details of a code change as input and a list of corresponding issues/code suggestions identified. Your goal is to inspect and review the issues/code suggestions, and score. \\

Be aware - the issues/suggestions may not always be correct or accurate, and you should evaluate them in relation to the actual code changes presented. \\

Carefully review both the suggestion content, and the related code change. Mistakes in the suggestions can occur. Make sure the suggestions are correct, and properly derived from the code changes. Mark each issue/suggestion as valid or invalid. \\

Score the suggestions: high scores (8 to 10) should be given to correct suggestions that address major bugs and issues, or security concerns. Lower scores (3 to 7) should be for correct suggestions addressing minor issues, code style, code readability, maintainability, etc. Don't give high scores to suggestions that are not crucial, and bring only small improvement or optimization.
Order the feedback the same way the suggestions are ordered in the input.
Response should be a valid YAML and nothing else.

\vspace{1em}
Here are the code changes and the corresponding code review suggestions.

=== Raw Source Control Code Change BEGIN === \\
code changes go here \\
=== Raw Source Control Code Change END ===

\vspace{1em}
=== Suggestions (YAML Format) BEGIN ===
\begin{verbatim}
```yaml
{
  "Function": ...,
  "Rationale": ...,
  "File": ...,
  "Line": ...,
  "Tag": ...
}
{
  "Function": ...,
  "Rationale": ...,
  "File": ...,
  "Line": ...,
  "Tag": ...
}
\end{verbatim}
=== Suggestions (YAML Format) END === \\

\vspace{1em}
Here is an example of the expected output format for reference. Please review and score each suggestion above.
\begin{verbatim}
=== Example Output Format ===
```yaml
{
"Suggestion_content": {rationale}
"Status": {valid/invalid}
"Sentiment": {positive if the suggestion is complimenting the 
             code changes, negative if the suggestion is 
             criticizing the code changes or neutral if uncear)}
"Line_matching": {yes/no, to check if the line number in the 
               issue description is matching the pre-pending 
               line number shown in the code change}
"Suggested score": {between 0-10}
"Score reason" : the code review suggestion is correct and 
               actionable because ...
}
```
\end{verbatim}

\vspace{1em}
Note: you should pay attention to the line number in the issue/code suggestion description, and judge if it is matching the pre-pending line number shown in the diff format. If the code suggestions are empty, just return an empty YAML string. \\

\vspace{1em}
Now begin. Please format your output according to the "Example Output Format" as shown above. Remember to do the review and score each code suggestion. 
\end{tcolorbox}

\newpage
\subsection{System Instruction for Issue Validator Agent}
\label{subsec:appendix_issue_validator_sys_instruct}
\begin{tcolorbox}[colback=white,  
                   colframe=black,    
                   coltext=blue,      
                   breakable,
                   fontupper=\fontfamily{pcr}\selectfont]  

Your input is a difference view of code changes, and a list of code issues for it. \\
Your goal is to inspect, review the issues and score each of them. \\
Be aware - the issues may not always be correct or accurate, and you should evaluate them in relation to the actual code changes presented. \\

\vspace{1em}
Review issues. Carefully review both the issue content, and the related code changes. Mistakes in the issue can occur. Make sure the issues are correct and properly derived from the code changes. Mark each issue as valid or invalid.

\vspace{1em}
Score issues. High scores (8 to 10) should be given to correct issues that address major bugs, or security concerns. Lower scores (3 to 7) should be for issues that address minor concerns for example code style, code readability, maintainability, etc. Don't give high scores to suggestions that are not crucial, and bring only small improvement or optimization to the code. Incorrect issues should be scored as 0. Order the feedback the same way the issues are ordered in the input. \\

\vspace{1em}
When reviewing issues, pay special attention to the following common mistakes:
\begin{itemize}[leftmargin=15pt]
    \item When reviewing documentation issues, verify the presence of existing docstrings. If a docstring is already present in any form (e.g., as a comment, a description, or a formal docstring), without nitpicking about its quality or completeness, mark the issue as invalid.
    \item When reviewing division by zero issues, carefully examine the code to verify that the variable in question is indeed used as a divisor in a division operation. If the variable is not involved in a division or if zero validation has already been performed, mark the issue as invalid.
\end{itemize}
Record verbosely your thought process in the output field "motivation".
\end{tcolorbox}

\newpage
\section{Appendix: Training Details}
\subsection{SFT Training Details}
\label{sec: sft_details}
Starting from llama-3.1-70b-instruct checkpoint, we fine-tune the 5000 SFT CQS examples using a sequence length of 8192 and a batch size of 64. The learning rate is set to 2e-7, with a warmup period of 10 steps and minimum rate of 0.1 with cosine annealing schedule. We only SFT for 1 epoch from the instruct checkpoint as we observe the performance start to degrade quickly after 1 epoch of training. 
We use 4x8 H100 GPUs with tensor parallelism set to 8. The default generation uses greedy decoding.

\subsection{Judge Model Training}
\label{sec: judge_training_details}
Starting from llama-3.1-70b-instruct checkpoint, we fine-tune the 1.6k critique examples using a sequence length of 8192 and a batch size of 32. The LR settings are identical to SFT training above. We observe the best accuracy at 4th epoch ($\approx$ 200 steps). 

\subsection{DPO Training Details}
\label{sec: dpo_details}
DPO training starts from the SFT-ed 70b checkpoints, we perform DPO training on the 8k preference pairs generated from judge results. Experiment setups follow a learning rate of  2e-7, with a warmup period of 25 steps and minimum rate of 0.5 with cosine schedule. The training employs a sequence length of 8192 and a batch size of 32, the best eval result is observed at 400 step. The DPO training is enabled with a \( \beta \)=0.1 and utilizes a sigmoid loss function. 

\section{Judge Model Study}
\label{sec: judge_study}
The eval set is randomly sampled thumb up/down human feedbacks, such signal can be noisy and may not clearly demonstrate the effectiveness of judge model. To further study the impact of judge training, we randomly sampled 1000 CQS issues from eval set, there exists 38 issues where the trained judge model disagreed with the base Llama 3.1-70b-instruct model -- which is, one model produces -1 score whereas the other produces positive score, or the opposite. We perform manual inspects on these cases, and the conclusion is that the trained judge model score is more aligned with developer sentiments: 35 out of 38 cases where they disagree, Judge model produces "-1" score  -- meaning the judge thinks this CQS issue is invalid, while base Llama model thinks it's a valid issue. In other words, besides clear lift on eval(62\% vs 56\%), the SFT training has nudged the model to produce more negative feedbacks on issues which developers complains. To further examine this effect, we compare the judge model’s scores on top 3 tags which received most complaints from beta-testing group (developers complained these issues associated with these 3 tags are typically low quality, as an example, the model is overly picky on incomplete docstring of a function/class). 
\begin{table}[h]
\centering
\caption{Comparing Jude Scores of Top User Complaint Tags}
\begin{tabular}{|l|l|l|l|}
\hline
\textbf{Judge Model} & \textbf{Documentation} & \textbf{UseConstant} & \textbf{DedupeLogic} \\
\hline
SFT Judge & Avg: 6.01, Std: 1.67 & Avg: 5.05, Std: 2.77 & Avg: 6.35, Std: 2.40 \\
\hline
Llama 3.1 & Avg: 6.28, Std: 1.52 & Avg: 5.52, Std: 2.54 & Avg: 6.58, Std: 2.18 \\
\hline
\end{tabular}
\end{table}

As evident in the table, after training with user feedback, score of most frequent complaint issues are suppressed.

\section{Failure Cases}
\label{sec:failure_cases}
Effective code review demands a profound comprehension of the code being reviewed, as well as extensive knowledge of both the programming language and the underlying business logic. Unfortunately, even SOTA LLMs has proven insufficient in meeting this requirement. The issues they identify are often misleading or trivial, as illustrated by the following two examples:

\begin{itemize}
    \item Missing language context

\begin{tcolorbox}[colback=white,  
                   colframe=blue,    
                   coltext=red,      
                   breakable,
                   fontupper=\fontfamily{pcr}\selectfont]

\textbf{C++ code}:
\begin{verbatim}
buffers = new(buffer.data()) EndToEndBuffers
\end{verbatim}
\vspace{1em}

\textbf{Issue Description}: The constructor now allocates memory for `EndToEndBuffers' using the new operator. Is it a good idea to use a smart pointer instead to avoid potential memory leaks?

\end{tcolorbox}
This C++ code utilizes the placement new operator, which does not allocate new memory but rather constructs an object in-place. Unfortunately, the LLM overlooked this subtlety and incorrectly flagged a memory leak.

    \item Nitpicking on extracting shared code
\begin{tcolorbox}[colback=white,  
                   colframe=blue,    
                   coltext=red,      
                   breakable,
                   fontupper=\fontfamily{pcr}\selectfont]
\textbf{C++ code}:
\begin{verbatim}
const auto& container = getContainer(ad, adToAugment);
\end{verbatim}

\vspace{1em}
\textbf{Issue Description}: The call `getContainer(ad, adToAugment)` to obtain `container` is repeated in this function (lines 461-463), and similarly in `FuncA` (lines 473-475) and `FuncB` (lines 486-488). Would it be beneficial to extract this common logic, perhaps into a private helper function within `ClassABC`, to improve maintainability and reduce redundancy?

\end{tcolorbox}
This C++ code is essentially a straightforward, three-line fetch operation. It's puzzling that the LLM is suggesting extracting this simple fetching logic into a separate function for deduplication purposes.

\end{itemize}

\end{document}